# Two-Dimensional Gas of Massless Dirac Fermions in Graphene

K.S. Novoselov[1], A.K. Geim[1], S.V. Morozov[2], D. Jiang[1], M.I. Katsnelson[3], I.V. Grigorieva[1], S.V. Dubonos[2], A.A. Firsov[2]

[1]Manchester Centre for Mesoscience and Nanotechnology, University of Manchester, Manchester, M13 9PL, UK

[2]Institute for Microelectronics Technology, 142432, Chernogolovka, Russia

[3]Institute for Molecules and Materials, Radboud University of Nijmegen, Toernooiveld 1, 6525 ED Nijmegen, the Netherlands

Electronic properties of materials are commonly described by quasiparticles that behave as non-relativistic electrons with a finite mass and obey the Schrödinger equation. Here we report a condensed matter system where electron transport is essentially governed by the Dirac equation and charge carriers mimic relativistic particles with zero mass and an effective "speed of light" $c_*$ ≈$10^6$m/s. Our studies of graphene – a single atomic layer of carbon – have revealed a variety of unusual phenomena characteristic of two-dimensional (2D) Dirac fermions. In particular, we have observed that a) the integer quantum Hall effect in graphene is anomalous in that it occurs at half-integer filling factors; b) graphene's conductivity never falls below a minimum value corresponding to the conductance quantum $e^2/h$, even when carrier concentrations tend to zero; c) the cyclotron mass $m_c$ of massless carriers with energy $E$ in graphene is described by equation $E = m_c c_*^2$; and d) Shubnikov-de Haas oscillations in graphene exhibit a phase shift of π due to Berry's phase.

Graphene is a monolayer of carbon atoms packed into a dense honeycomb crystal structure that can be viewed as either an individual atomic plane extracted from graphite or unrolled single-wall carbon nanotubes or as a giant flat fullerene molecule. This material was not studied experimentally before and, until recently [1,2], presumed not to exist. To obtain graphene samples, we used the original procedures described in [1], which involve micromechanical cleavage of graphite followed by identification and selection of monolayers using a combination of optical, scanning-electron and atomic-force microscopies. The selected graphene films were further processed into multi-terminal devices such as the one shown in Fig. 1, following standard microfabrication procedures [2]. Despite being only one atom thick and unprotected from the environment, our graphene devices remain stable under ambient conditions and exhibit high mobility of charge carriers. Below we focus on the physics of "ideal" (single-layer) graphene which has a different electronic structure and exhibits properties qualitatively different from those characteristic of either ultra-thin graphite films (which are semimetals and whose material properties were studied recently [2-5]) or even of our other devices consisting of just two layers of graphene (see further).

Figure 1 shows the electric field effect [2-4] in graphene. Its conductivity σ increases linearly with increasing gate voltage $V_g$ for both polarities and the Hall effect changes its sign at $V_g$ ≈0. This behaviour shows that substantial concentrations of electrons (holes) are induced by positive (negative) gate voltages. Away from the transition region $V_g$ ≈0, Hall coefficient $R_H = 1/ne$ varies as $1/V_g$ where $n$ is the concentration of electrons or holes and $e$ the electron charge. The linear dependence $1/R_H \propto V_g$ yields $n = \alpha \cdot V_g$ with α ≈7.3·$10^{10}$cm$^{-2}$/V, in agreement with the theoretical estimate $n/V_g$ ≈7.2·$10^{10}$cm$^{-2}$/V for the surface charge density induced by the field effect (see Fig. 1's caption). The agreement indicates that all the induced carriers are mobile and there are no trapped charges in graphene. From the linear dependence σ($V_g$) we found carrier mobilities μ =σ/ne, which reached up to 5,000 cm$^2$/Vs for both electrons and holes, were independent of temperature $T$ between 10 and 100K and probably still limited by defects in parent graphite.

To characterise graphene further, we studied Shubnikov-de Haas oscillations (SdHO). Figure 2 shows examples of these oscillations for different magnetic fields $B$, gate voltages and temperatures. Unlike ultra-thin graphite [2], graphene exhibits only one set of SdHO for both electrons and holes. By using standard fan diagrams [2,3], we have determined the fundamental SdHO frequency $B_F$ for various $V_g$. The resulting dependence of $B_F$ as a function of $n$ is plotted in Fig. 3a. Both carriers exhibit the same linear dependence $B_F = \beta \cdot n$ with $\beta \approx 1.04 \cdot 10^{-15}$ T·m$^2$ (±2%). Theoretically, for any 2D system $\beta$ is defined only by its degeneracy $f$ so that $B_F = \phi_0 n/f$, where $\phi_0 = 4.14 \cdot 10^{-15}$ T·m$^2$ is the flux quantum. Comparison with the experiment yields $f = 4$, in agreement with the double-spin and double-valley degeneracy expected for graphene [6,7] (cf. caption of Fig. 2). Note however an anomalous feature of SdHO in graphene, which is their phase. In contrast to conventional metals, graphene's longitudinal resistance $\rho_{xx}(B)$ exhibits maxima rather than minima at integer values of the Landau filling factor $\nu$ (Fig. 2a). Fig. 3b emphasizes this fact by comparing the phase of SdHO in graphene with that in a thin graphite film [2]. The origin of the "odd" phase is explained below.

Another unusual feature of 2D transport in graphene clearly reveals itself in the $T$-dependence of SdHO (Fig. 2b). Indeed, with increasing $T$ the oscillations at high $V_g$ (high $n$) decay more rapidly. One can see that the last oscillation ($V_g \approx 100$V) becomes practically invisible already at 80K whereas the first one ($V_g < 10$V) clearly survives at 140K and, in fact, remains notable even at room temperature. To quantify this behaviour we measured the $T$-dependence of SdHO's amplitude at various gate voltages and magnetic fields. The results could be fitted accurately (Fig. 3c) by the standard expression $T/\sinh(2\pi^2 k_B T m_c/\hbar eB)$, which yielded $m_c$ varying between $\approx 0.02$ and $0.07 m_0$ ($m_0$ is the free electron mass). Changes in $m_c$ are well described by a square-root dependence $m_c \propto n^{1/2}$ (Fig. 3d).

To explain the observed behaviour of $m_c$, we refer to the semiclassical expressions $B_F = (\hbar/2\pi e)S(E)$ and $m_c = (\hbar^2/2\pi)\partial S(E)/\partial E$ where $S(E) = \pi k^2$ is the area in k-space of the orbits at the Fermi energy $E(k)$ [8]. Combining these expressions with the experimentally-found dependences $m_c \propto n^{1/2}$ and $B_F = (h/4e)n$ it is straightforward to show that $S$ must be proportional to $E^2$ which yields $E \propto k$. Hence, the data in Fig. 3 unambiguously prove the linear dispersion $E = \hbar k c_*$ for both electrons and holes with a common origin at $E = 0$ [6,7]. Furthermore, the above equations also imply $m_c = E/c_*^2 = (h^2n/4\pi c_*^2)^{1/2}$ and the best fit to our data yields $c_* \approx 1 \cdot 10^6$ m/s, in agreement with band structure calculations [6,7]. The employed semiclassical model is fully justified by a recent theory for graphene [9], which shows that SdHO's amplitude can indeed be described by the above expression $T/\sinh(2\pi^2 k_B T m_c/\hbar eB)$ with $m_c = E/c_*^2$. Note that, even though the linear spectrum of fermions in graphene (Fig. 3e) implies zero rest mass, their cyclotron mass is not zero.

The unusual response of massless fermions to magnetic field is highlighted further by their behaviour in the high-field limit where SdHO evolve into the quantum Hall effect (QHE). Figure 4 shows Hall conductivity $\sigma_{xy}$ of graphene plotted as a function of electron and hole concentrations in a constant field $B$. Pronounced QHE plateaux are clearly seen but, surprisingly, they do not occur in the expected sequence $\sigma_{xy} = (4e^2/h)N$ where $N$ is integer. On the contrary, the plateaux correspond to half-integer $\nu$ so that the first plateau occurs at $2e^2/h$ and the sequence is $(4e^2/h)(N + ½)$. Note that the transition from the lowest hole ($\nu = -½$) to lowest electron ($\nu = +½$) Landau level (LL) in graphene requires the same number of carriers ($\Delta n = 4B/\phi_0 \approx 1.2 \cdot 10^{12}$ cm$^{-2}$) as the transition between other nearest levels (cf. distances between minima in $\rho_{xx}$). This results in a ladder of equidistant steps in $\sigma_{xy}$ which are not interrupted when passing through zero. To emphasize this highly unusual behaviour, Fig. 4 also shows $\sigma_{xy}$ for a graphite film consisting of only two graphene layers where the sequence of plateaux returns to normal and the first plateau is at $4e^2/h$, as in the conventional QHE. We attribute this qualitative transition between graphene and its two-layer counterpart to the fact that fermions in the latter exhibit a finite mass near $n \approx 0$ (as found experimentally; to be published elsewhere) and can no longer be described as massless Dirac particles.

The half-integer QHE in graphene has recently been suggested by two theory groups [10,11], stimulated by our work on thin graphite films [2] but unaware of the present experiment. The effect is single-particle and intimately related to subtle properties of massless Dirac fermions, in particular, to the existence of both electron- and hole-like Landau states at exactly zero energy [9-12]. The latter can be viewed as a direct consequence of the Atiyah-Singer index theorem that plays an important role in quantum field theory and the theory of superstrings [13,14]. For the case of 2D massless Dirac fermions, the theorem guarantees the existence of Landau states at $E=0$ by relating the difference in the number of such states with opposite chiralities to the total flux through the system (note that magnetic field can also be inhomogeneous).

To explain the half-integer QHE qualitatively, we invoke the formal expression [9-12] for the energy of massless relativistic fermions in quantized fields, $E_N = [2e\hbar c_*^2 B(N + \frac{1}{2} \pm \frac{1}{2})]^{1/2}$. In QED, sign ± describes two spins whereas in the case of graphene it refers to "pseudospins". The latter have nothing to do with the real spin but are "built in" the Dirac-like spectrum of graphene, and their origin can be traced to the presence of two carbon sublattices. The above formula shows that the lowest LL ($N = 0$) appears at $E = 0$ (in agreement with the index theorem) and accommodates fermions with only one (minus) projection of the pseudospin. All other levels $N \geq 1$ are occupied by fermions with both (±) pseudospins. This implies that for $N = 0$ the degeneracy is half of that for any other $N$. Alternatively, one can say that all LL have the same "compound" degeneracy but zero-energy LL is shared equally by electrons and holes. As a result the first Hall plateau occurs at half the normal filling and, oddly, both $\nu = -\frac{1}{2}$ and $+\frac{1}{2}$ correspond to the same LL ($N = 0$). All other levels have normal degeneracy $4B/\phi_0$ and, therefore, remain shifted by the same ½ from the standard sequence. This explains the QHE at $\nu = N + \frac{1}{2}$ and, at the same time, the "odd" phase of SdHO (minima in $\rho_{xx}$ correspond to plateaux in $\rho_{xy}$ and, hence, occur at half-integer $\nu$; see Figs. 2&3), in agreement with theory [9-12]. Note however that from another perspective the phase shift can be viewed as the direct manifestation of Berry's phase acquired by Dirac fermions moving in magnetic field [15,16].

Finally, we return to zero-field behaviour and discuss another feature related to graphene's relativistic-like spectrum. The spectrum implies vanishing concentrations of both carriers near the Dirac point $E = 0$ (Fig. 3e), which suggests that low-$T$ resistivity of the zero-gap semiconductor should diverge at $V_g \approx 0$. However, neither of our devices showed such behaviour. On the contrary, in the transition region between holes and electrons graphene's conductivity never falls below a well-defined value, practically independent of $T$ between 4 and 100K. Fig. 1c plots values of the maximum resistivity $\rho_{max}(B = 0)$ found in 15 different devices, which within an experimental error of ≈15% all exhibit $\rho_{max} \approx 6.5$k$\Omega$, independent of their mobility that varies by a factor of 10. Given the quadruple degeneracy $f$, it is obvious to associate $\rho_{max}$ with $h/fe^2 = 6.45$k$\Omega$ where $h/e^2$ is the resistance quantum. We emphasize that it is the *resistivity* (or *conductivity*) rather than resistance (or conductance), which is quantized in graphene (i.e., resistance $R$ measured experimentally was not quantized but scaled in the usual manner as $R = \rho L/w$ with changing length $L$ and width $w$ of our devices). Thus, the effect is completely different from the conductance quantization observed previously in quantum transport experiments.

However surprising, the minimum conductivity is an intrinsic property of electronic systems described by the Dirac equation [17-20]. It is due to the fact that, in the presence of disorder, localization effects in such systems are strongly suppressed and emerge only at exponentially large length scales. Assuming the absence of localization, the observed minimum conductivity can be explained qualitatively by invoking Mott's argument [21] that mean-free-path $l$ of charge carriers in a metal can never be shorter that their wavelength $\lambda_F$. Then, $\sigma = ne\mu$ can be re-written as $\sigma = (e^2/h)k_F l$ and, hence, $\sigma$ cannot be smaller than $\approx e^2/h$ per each type of carriers. This argument is known to have failed for 2D systems with a parabolic spectrum where disorder leads to localization and eventually to insulating behaviour [17,18]. For the case of 2D Dirac fermions, no localization is expected [17-20] and, accordingly, Mott's argument can be used. Although there is a broad theoretical consensus [18-23,10,11] that a 2D gas of Dirac fermions should exhibit a minimum conductivity of about $e^2/h$, this quantization was not expected to be accurate and most theories suggest a value of $\approx e^2/\pi h$, in disagreement with the experiment.

In conclusion, graphene exhibits electronic properties distinctive for a 2D gas of particles described by the Dirac rather than Schrödinger equation. This 2D system is not only interesting in itself but also allows one to access – in a condensed matter experiment – the subtle and rich physics of quantum electrodynamics [24-27] and provides a bench-top setting for studies of phenomena relevant to cosmology and astrophysics [27,28].

1. Novoselov, K.S. *et al. PNAS* **102**, 10451 (2005).
2. Novoselov, K.S. *et al. Science* **306**, 666 (2004); cond-mat/0505319.
3. Zhang, Y., Small, J.P., Amori, M.E.S. & Kim, P. *Phys. Rev. Lett.* **94**, 176803 (2005).
4. Berger, C. *et al. J. Phys. Chem. B*, **108**, 19912 (2004).
5. Bunch, J.S., Yaish, Y., Brink, M., Bolotin, K. & McEuen, P.L. *Nanoletters* 5, 287 (2005).
6. Dresselhaus, M.S. & Dresselhaus, G. *Adv. Phys*. **51**, 1 (2002).
7. Brandt, N.B., Chudinov, S.M. & Ponomarev, Y.G. *Semimetals 1: Graphite and Its Compounds* (North-Holland, Amsterdam, 1988).
8. Vonsovsky, S.V. and Katsnelson, M.I. *Quantum Solid State Physics* (Springer, New York, 1989).
9. Gusynin, V.P. & Sharapov, S.G. *Phys. Rev. B* **71**, 125124 (2005).
10. Gusynin, V.P. & Sharapov, S.G. cond-mat/0506575.
11. Peres, N.M.R., Guinea, F. & Castro Neto, A.H. cond-mat/0506709.
12. Zheng, Y. & Ando, T. *Phys. Rev. B* **65**, 245420 (2002).
13. Kaku, M. *Introduction to Superstrings* (Springer, New York, 1988).
14. Nakahara, M. *Geometry, Topology and Physics* (IOP Publishing, Bristol, 1990).
15. Mikitik, G. P. & Sharlai, Yu.V. *Phys. Rev. Lett.* **82**, 2147 (1999).
16. Luk'yanchuk, I.A. & Kopelevich, Y. *Phys. Rev. Lett.* **93**, 166402 (2004).
17. Abrahams, E., Anderson, P.W., Licciardello, D.C. & Ramakrishnan, T.V. *Phys. Rev. Lett.* 42, 673 (1979).
18. Fradkin, E. *Phys. Rev.* B **33**, 3263 (1986).
19. Lee, P.A. *Phys. Rev. Lett.* **71**, 1887 (1993).
20. Ziegler, K. *Phys. Rev. Lett.* **80**, 3113 (1998).
21. Mott, N.F. & Davis, E.A. *Electron Processes in Non-Crystalline Materials* (Clarendon Press, Oxford, 1979).
22. Morita, Y. & Hatsugai, Y. *Phys. Rev. Lett.* **79**, 3728 (1997).
23. Nersesyan, A.A., Tsvelik, A.M. & Wenger, F. *Phys. Rev. Lett.* **72**, 2628 (1997).
24. Rose, M.E. *Relativistic Electron Theory* (John Wiley, New York, 1961).
25. Berestetskii, V.B., Lifshitz, E.M. & Pitaevskii, L.P. *Relativistic Quantum Theory* (Pergamon Press, Oxford, 1971).
26. Lai, D. *Rev. Mod. Phys.* **73**, 629 (2001).
27. Fradkin, E. *Field Theories of Condensed Matter Systems* (Westview Press, Oxford, 1997).
28. Volovik, G.E. *The Universe in a Helium Droplet* (Clarendon Press, Oxford, 2003).

**Acknowledgements** This research was supported by the EPSRC (UK). We are most grateful to L. Glazman, V. Falko, S. Sharapov and A. Castro Netto for helpful discussions. K.S.N. was supported by Leverhulme Trust. S.V.M., S.V.D. and A.A.F. acknowledge support from the Russian Academy of Science and INTAS.

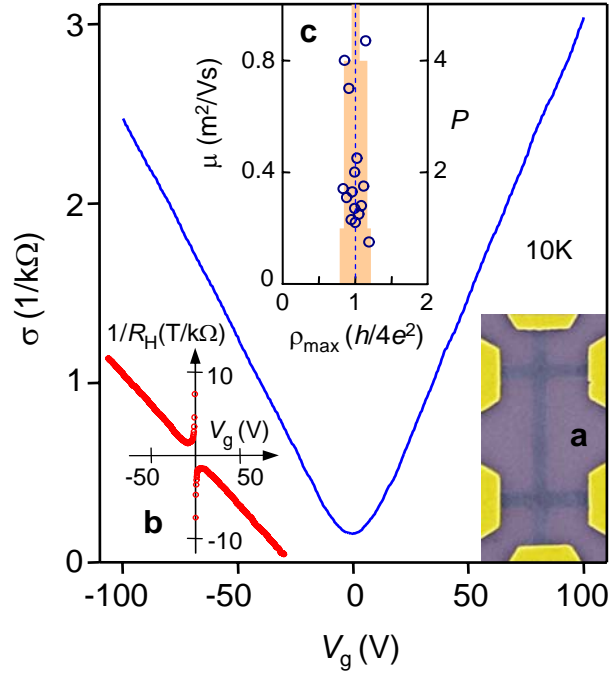

Figure 1. Electric field effect in graphene. **a**, Scanning electron microscope image of one of our experimental devices (width of the central wire is 0.2μm). False colours are chosen to match real colours as seen in an optical microscope for larger areas of the same materials. Changes in graphene's conductivity σ (main panel) and Hall coefficient $R_H$ (**b**) as a function of gate voltage $V_g$. σ and $R_H$ were measured in magnetic fields $B$ =0 and 2T, respectively. The induced carrier concentrations $n$ are described by [2] $n/V_g = \varepsilon_0\varepsilon/te$ where $\varepsilon_0$ and ε are permittivities of free space and $SiO_2$, respectively, and $t \approx$300 nm is the thickness of $SiO_2$ on top of the Si wafer used as a substrate. $R_H = 1/ne$ is inverted to emphasize the linear dependence $n \propto V_g$. $1/R_H$ diverges at small $n$ because the Hall effect changes its sign around $V_g$ =0 indicating a transition between electrons and holes. Note that the transition region ($R_H \approx 0$) was often shifted from zero $V_g$ due to chemical doping [2] but annealing of our devices in vacuum normally allowed us to eliminate the shift. The extrapolation of the linear slopes σ($V_g$) for electrons and holes results in their intersection at a value of σ indistinguishable from zero. **c**, Maximum values of resistivity ρ =1/σ (circles) exhibited by devices with different mobilites μ (left y-axis). The histogram (orange background) shows the number $P$ of devices exhibiting $\rho_{max}$ within 10% intervals around the average value of $\approx h/4e^2$. Several of the devices shown were made from 2 or 3 layers of graphene indicating that the quantized minimum conductivity is a robust effect and does not require "ideal" graphene.

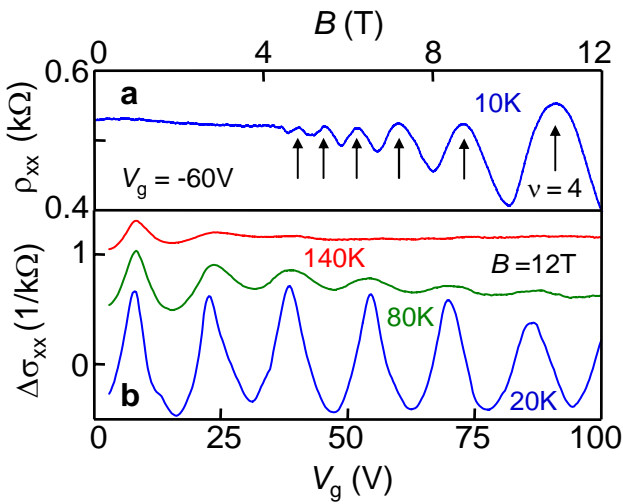

Figure 2. Quantum oscillations in graphene. SdHO at constant gate voltage $V_g$ as a function of magnetic field $B$ (**a**) and at constant $B$ as a function of $V_g$ (**b**). Because μ does not change much with $V_g$, the constant-$B$ measurements (at a constant $\omega_c\tau =\mu B$) were found more informative. Panel **b** illustrates that SdHO in graphene are more sensitive to $T$ at high carrier concentrations. The $\Delta\sigma_{xx}$-curves were obtained by subtracting a smooth (nearly linear) increase in σ with increasing $V_g$ and are shifted for clarity. SdHO periodicity $\Delta V_g$ in a constant $B$ is determined by the density of states at each Landau level ($\alpha\Delta V_g = fB/\phi_0$) which for the observed periodicity of ≈15.8V at $B$ =12T yields a quadruple degeneracy. Arrows in **a** indicate integer ν (e.g., ν =4 corresponds to 10.9T) as found from SdHO frequency $B_F \approx$43.5T. Note the absence of any significant contribution of universal conductance fluctuations (see also Fig. 1) and weak localization magnetoresistance, which are normally intrinsic for 2D materials with so high resistivity.

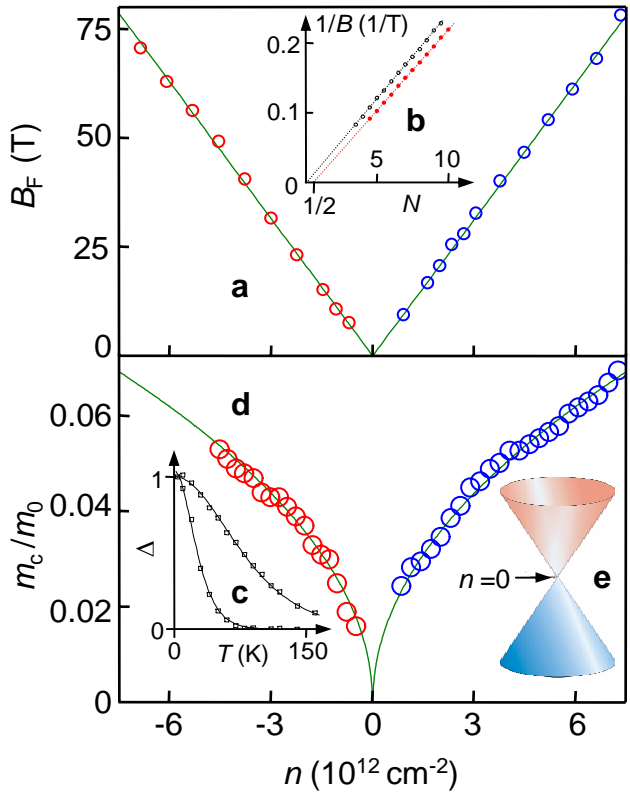

Figure 3. Dirac fermions of graphene. **a**, Dependence of $B_F$ on carrier concentration $n$ (positive $n$ correspond to electrons; negative to holes). **b**, Examples of fan diagrams used in our analysis [2] to find $B_F$. $N$ is the number associated with different minima of oscillations. Lower and upper curves are for graphene (sample of Fig. 2a) and a 5-nm-thick film of graphite with a similar value of $B_F$, respectively. Note that the curves extrapolate to different origins; namely, to $N = \frac{1}{2}$ and 0. In graphene, curves for all $n$ extrapolate to $N = \frac{1}{2}$ (cf. [2]). This indicates a phase shift of $\pi$ with respect to the conventional Landau quantization in metals. The shift is due to Berry's phase [9,15]. **c**, Examples of the behaviour of SdHO amplitude $\Delta$ (symbols) as a function of $T$ for $m_c \approx 0.069$ and $0.023m_0$; solid curves are best fits. **d**, Cyclotron mass $m_c$ of electrons and holes as a function of their concentration. Symbols are experimental data, solid curves the best fit to theory. **e**, Electronic spectrum of graphene, as inferred experimentally and in agreement with theory. This is the spectrum of a zero-gap 2D semiconductor that describes massless Dirac fermions with $c_*$ 300 times less than the speed of light.

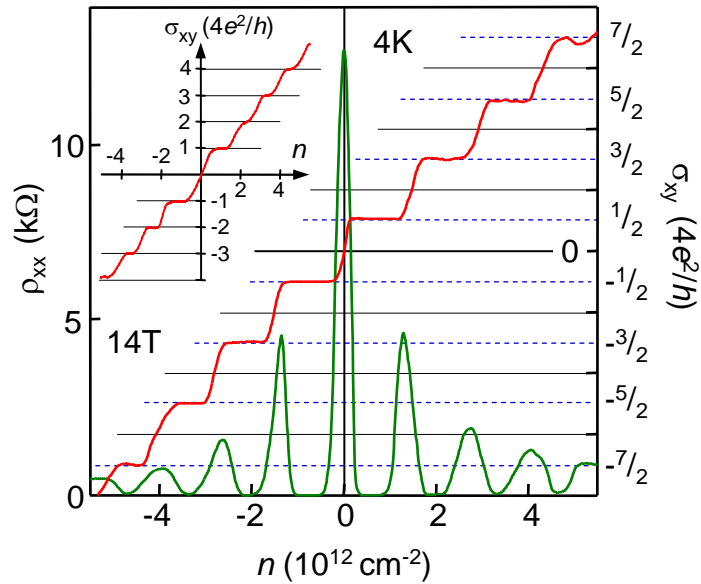

Figure 4. Quantum Hall effect for massless Dirac fermions. Hall conductivity $\sigma_{xy}$ and longitudinal resistivity $\rho_{xx}$ of graphene as a function of their concentration at $B = 14$T. $\sigma_{xy} = (4e^2/h)\nu$ is calculated from the measured dependences of $\rho_{xy}(V_g)$ and $\rho_{xx}(V_g)$ as $\sigma_{xy} = \rho_{xy}/(\rho_{xy} + \rho_{xx})^2$. The behaviour of $1/\rho_{xy}$ is similar but exhibits a discontinuity at $V_g \approx 0$, which is avoided by plotting $\sigma_{xy}$. Inset: $\sigma_{xy}$ in "two-layer graphene" where the quantization sequence is normal and occurs at integer $\nu$. The latter shows that the half-integer QHE is exclusive to "ideal" graphene.